\documentclass[prd,aps,floats,nofootinbib,tightenlines,showpacs]{revtex4}
\usepackage{epsfig,epsf}
\usepackage{amssymb}
\usepackage{amsmath}
\usepackage{graphicx}
\usepackage{latexsym}
\usepackage{slashed}
\usepackage[normalem]{ulem}
\usepackage{pslatex,bm}

\def\){\right)}
\def\({\left(}
\def\]{\right]}
\def\[{\left[}

\def\psibar{\overline{\psi}}

\def\multsp{ }

\newcommand{\mcal}[1]{{\mathcal #1}}
\newcommand{\be}{\begin{equation}}
\newcommand{\ee}{\end{equation}}
\newcommand{\gsim}{\, \raisebox{-0.8ex}{$\stackrel{\textstyle >}{\sim}$ }}
\newcommand{\lsim}{\, \, \raisebox{-0.8ex}{$\stackrel{\textstyle <}{\sim}$ }}

\newcommand{\roughly}[1]%
{\mathrel{\raise.4ex\hbox{$#1$\kern-.75em\lower1ex\hbox{$\sim$}}}}

\newcommand\beq{\begin{eqnarray}}\newcommand\eeq{\end{eqnarray}}

\def\Dsl{\,\raise.15ex \hbox{/}\mkern-12.8mu D}

\def\fm3{fm$^{-3}$}

\begin{document}
%\begin{frontmatter}
%
\preprint{\vbox{\hbox{LA-UR-04-3596}}}

\bigskip
\bigskip

\title{Phase Structure of 2-Flavor Quark Matter: Heterogeneous Superconductors}

\author{Sanjay Reddy and Gautam Rupak}

\affiliation{Theoretical Division, Los Alamos National Laboratory, Los
Alamos, NM 87545 \\ 
}
\begin{abstract}
We analyze the free energy of charge and color neutral 2-flavor quark
matter within the BCS approximation. We consider both the homogeneous
gapless superconducting phase and the heterogeneous mixed phase where
normal and BCS superconducting phases coexist. We calculate the
surface tension between normal and superconducting phases and use it
to compare the free energies of the gapless and mixed phases. Our
calculation, which retains only the leading order gradient
contribution to the free energy, indicates that the mixed phase is
energetically favored over an interesting range of densities of
relevance to 2 flavor quark matter in neutron stars. 
\end{abstract}
%25.75Nq  quark deconfinement, QGP, and phase transitions
%26.60+c  nuclear matter aspects of neutron stars
%97.60.Jd neutron stars
\maketitle

\section{Introduction}
At high baryon density, we expect on general grounds that nuclear
matter will undergo a phase transition to the de-confined quark phase.
Although the nature of this transition and its location remain
unclear, theory suggests that the quark phase is likely to be a color
superconductor. This expectation is several decades old
\cite{Barrois:1977xd,Bailin:1984bm}. Recent work based on effective
models for the quark-quark interaction predict a superconducting gap
$\Delta \simeq 100$ MeV for a quark chemical potential $\mu \simeq
400$ MeV \cite{Alford:1998zt,Rapp:1998zu} and has generated renewed
interest in the field. The pairing or BCS (Bardeen-Cooper- Schrieffer)
instability is strongest in the spin-zero, color and flavor
antisymmetric channel, naturally resulting in pairing between quarks of
different flavors. Any source of flavor symmetry breaking will induce
a stress on the paired state since it will act to move the Fermi
surfaces of the different flavors apart. There has been much recent
interest in how this will affect the ground state properties of quark
matter. In this work we investigate the role of the large isospin
breaking induced by electromagnetism in 2 flavor quark matter.

In the absence of any flavor symmetry breaking the ground state of
bulk three flavor quark matter is expected to be the
Color-Flavor-Locked state \cite{Alford:1998mk}.  In this phase, all
nine quarks participate in pairing and the ${ SU(3)_{\rm color}}
\times SU(3)_L \times SU(3)_R \times U(1)_B$ symmetry of QCD is broken
down to the global diagonal $SU(3)$ symmetry. Chiral symmetry breaking
results in light pseudo-Goldstone modes with the quantum numbers of
the pions, kaons and eta's.  This phase is neutral with respect to
electric and color charge as it contains equal numbers of up, down and
strange quarks.  A finite strange quark mass breaks flavor
symmetry. For small values of the strange quark mass ($m_s \ll \Delta$
) the stress on the CFL state is resolved via a novel mechanism
wherein the light pseudo-Goldstone modes condense
\cite{Bedaque:2001je,Kaplan:2001qk}.  Bedaque and Schafer have shown
that the stress induced by the strange quark mass can result in the
condensation of neutral kaons in the CFL state \cite{Bedaque:2001je}. At intermediate values of the strange quark mass corresponding to
$m_s^2/\mu \simeq \Delta$, the situation is less clear. Recent,
calculations suggest the possible existence of a color-flavor locked
phase with non-trivial gapless quark excitations \cite{Alford:2003fq}.

In this article we study the case where the (effective) strange quark
mass is large compared to the quark chemical potential. Here no
strange quarks are present and the excess charge of the up and down
quarks is neutralized by electrons, which in turn induces a splitting
$\delta\mu=(\mu_u-\mu_d)/2$ between the up and down quark chemical
potentials. If $\delta\mu$ is large compared to the pairing energy,
the pairing breaks down and the normal state is favored. At
intermediate values, interesting new phases are possible and have been
previously suggested. This includes the homogeneous gapless
superconducting phase \cite{Shovkovy:2003uu} . The gapless
superconducting phase, also called the Sarma or breached
phase\cite{Sarma:1963}, was found to be an unstable maxima of the free energy in electron superconductors. In  the quark matter context, a meta-stable gapless state was first discussed in  
Ref.~\cite{Alford:1999xc}.  In the condensed matter context, stable gapless phases were discussed by Liu and Wilczek \cite{Liu:2002gi}. Subsequently, Shovkovy and Huang  showed that charge neutrality could stabilize the gapless phase in quark matter\cite{Shovkovy:2003uu}. The other possibility is 
the heterogeneous mixed phase where normal and BCS superconducting
phases coexist \cite{Bedaque:1999nu,Bedaque:2003hi}. Although we do not
study it in this work, we note that there exists a small interval in
$\delta \mu/\Delta$ where crystalline superconductivity could be
favored \cite{Alford:2000ze}. This crystalline phase, which is also
called the Larkin, Ovchinnikov, Fulde and Ferrel (LOFF) phase, is
characterized by a spatial variation of $\Delta$ on a scale $1/\delta
\mu$.  The mixed phase that we consider here shares some features with
the LOFF phase: spatial variation of $\Delta$ and crystalline
structure.  In the mixed phase electric charge neutrality is satisfied
as a global constraint- a positively charged BCS phase coexists and
neutralizes the negatively charged normal phase. As emphasized in
earlier work by Glendenning, the mixed phase is a generic possibility
associated with first order transitions involving two chemical
potentials \cite{Glendenning:1992vb}. This charge separation leads to
long-range order and crystalline structure at low temperature
\cite{Glendenning:1995rd}. The description and competition between the
mixed and the gapless phases is the subject of this article. It has
been shown that mixed phase is energetically favored only for small
values of the surface tension \cite{Shovkovy:2003ce}.  In this work,
we calculate the surface tension for the first time and show that it
is indeed small. Consequently we find that the heterogeneous mixed
phase is favored.

We will begin by deriving the free energy of color and charge neutral
matter containing light (up and down) quarks and electrons in
Sec. \ref{free}. In concordance with earlier findings we show the
existence of two stable and neutral bulk ground states: the
homogeneous gapless state and the heterogeneous mixed phase. In
Sec. \ref{surface}, we calculate the surface tension between the
normal and superconducting phases in the mixed phase. We will then use
these results to compare the free energies of the homogeneous and
heterogeneous superconducting phases in Sec.\ref{phase}. Finally in
Sec. \ref{discuss} we conclude with a discussion of our results and
explore some of its consequences for astrophysics of neutron stars
that may contain quark matter. 

%==============================================
\section{Free Energy}
\label{free}
%==============================================
We consider a system of massless electrons, and two degenerate light
quarks (up and down), with three color degrees of freedom at finite
quark chemical potential $\hat{\mu}= \mu+\mu_Q~Q+\mu_{8}~T_8 +
\mu_3~T_3$.  We will assume that the $\langle \bar{q}q\rangle$
condensation for light quarks is small in the quark phase and will
restrict our investigation to quark masses that are small compared to
the chemical potential. We have written the quark chemical potential
in terms of the baryon chemical potential ($\mu=\mu_B/3$), the
electric charge chemical potential $\mu_Q$ and the color chemical
potentials $\mu_3$ and $\mu_8$. The electric charge matrix and the
color matrices are defined below.
\begin{align}
Q=\(\begin{array}{cc}\frac{2}{3}&0\\0&-\frac{1}{3}\end{array}\)
=\frac{1}{6}+\frac{1}{2}\tau_3 \,, \quad 
T_{3}\equiv2\lambda_3
=\(\begin{array}{ccc}1&0&0\\0&-1&0\\
0&0&0\end{array}\)\,, \quad
T_{8}\equiv2\sqrt{3}\lambda_8
=\(\begin{array}{ccc}1&0&0\\0&1&0\\
0&0&-2\end{array}\)\,.
%\Rightarrow
%\mu=&\(\mu_b+\mu_c+\frac{\mu_Q}{6}\)+\frac{\mu_Q}{2}\tau_3+2\mu_3\lambda_3+
%2\sqrt{3}\mu_8\lambda_8\equiv
%\mu_0+\delta\mu\tau_3+2\mu_3\lambda_3+
%2\sqrt{3}\mu_8\lambda_8.\notag
\end{align}
It is convenient to write the quark chemical potential (matrix) as 
\begin{equation} 
\hat{\mu} = \mu_0 + \delta\mu~\tau_3 + \mu_3~T_3 + \mu_8~T_8,
\end{equation} 
where $\mu_0=\mu + \mu_Q/6$ and $\delta\mu=\mu_Q/2$.  The Pauli
matrices $\tau_i$'s act in flavor space and the Gell-Mann matrices
$\lambda_i$'s act in color space.

We model the pairing interaction by a four-quark operator with the
quantum numbers of the single gluon exchange interaction. As
Eq.~(\ref{oge}) shows, one gluon exchange is attractive in the color
anti-symmetric channel and repulsive in the symmetric channel.
\begin{align}
\label{oge}
-(\psibar\gamma_\mu \lambda_A\psi)(\psibar\gamma^\mu\lambda_A\psi)
=(\psibar_\alpha\gamma_\mu\psi_\beta)&
(\psibar_\gamma\gamma^\mu\psi_\delta)\[ 
\frac{1}{3}({{\delta }_
     {\alpha \multsp \beta }}{{\delta }_
      {\gamma \multsp \delta }}-
     {{\delta }_{\alpha \multsp \delta }}
     {{\delta }_{\beta \multsp \gamma }}
      )
-\frac{1}{6}
       ({{\delta }_{\alpha \multsp \beta }}
        {{\delta }_{\gamma \multsp \delta }}
         +{{\delta }_{\alpha \multsp \delta }}
          {{\delta }_{\beta \multsp \gamma }})
\],
\end{align}  
where the first term corresponds to the attractive color antisymmetric
$\overline{\bm 3}$ channel. For pairing, we only consider the attractive
channel.  Further, we consider the pairing to be in the spin zero,
$S$-wave and flavor antisymmetric channel. So, the four-quark operator
is taken to be
\begin{align}
g(\psibar_\alpha\gamma_5\tau_2\psi^c_\beta)(\psibar^c_\gamma\gamma_5\tau_2\psi_\delta) ({{\delta }_{\alpha \multsp \gamma }}
  {{\delta }_{\beta \multsp \delta }}
   -{{\delta }_{\alpha \multsp \delta }}
    {{\delta }_{\beta \multsp \gamma }}),\\
\psi^c=C\psibar^T=i\gamma_2\psi^{*},
\ \ \psibar^c=\psi^T C=i\psi^T\gamma_2\gamma_0. \notag
\end{align}
Using the 2SC ansatz 
$\psibar^c_\alpha\gamma_5\tau_2\psi_\beta=
i\Delta\epsilon_{\alpha\beta 3}/(4 g)$ (where blue quarks remain unpaired) 
 for the BCS gap $\Delta$ in mean field approximation, we get for the quark
Lagrangian 
\begin{align}
\mcal L+\mu N=&\psibar\(i\slashed{\partial}+\gamma_0\mu-m\)\psi
+g(\psibar_\alpha\gamma_5\tau_2\psi^c_\beta)
(\psibar^c_\gamma\gamma_5\tau_2\psi_\delta) ({{\delta }_{\alpha
\multsp \gamma }} {{\delta }_{\beta \multsp \delta }} -{{\delta
}_{\alpha \multsp \delta }} {{\delta }_{\beta \multsp \gamma }})\\
=&\psibar^{(b)}
\(i\slashed{\partial}+\gamma_0\mu_b\)\psi^{(b)}-\frac{\Delta^2}{4 g}
-\frac{1}{2}~\overline\Psi^{(rg)}_i M_{i j}\Psi^{(rg)}_j, \notag \\
\mu_b=&\mu_0-2\mu_8+\delta\mu~\tau_3,\notag\\
\overline\Psi^{(rg)}=&(\begin{array}{cc}\psibar&\psibar^c\end{array}),\notag\\
M=&\(\begin{array}{cc}
i\slashed{\partial}-m+\gamma_0(\mu_0+\delta\mu\tau_3+\mu_3\sigma_3+\mu_8)&
-\gamma_5\tau_2\sigma_2\Delta\\
\gamma_5\tau_2\sigma_2\Delta&i\slashed{\partial}-m
-\gamma_0(\mu_0+\delta\mu\tau_3+\mu_3\sigma_3+\mu_8)
\end{array}\).\notag
\end{align}
The Dirac spinor fields $\psi^{(b)}$, $\Psi^{(rg)}$ represent the
blue, and red-green quark fields respectively. The Pauli matrices
$\tau_i$'s act in flavor space, and $\sigma_i$'s act in the red-green
color space. Integrating over the fermionic variables in the partition
function, the free energy for a constant gap field $\Delta$ is given
by
\begin{align}\label{freeenergy}
\Omega_{\mathrm{2SC}}(\Delta)=&\frac{\Delta^2}{4 g}
+
2 \sum_{\pm}\int \frac{{d^3}p}{{{(2\multsp \pi )}^3}}
  ({\sqrt{{\bm p^2}+{m^2}}}
   -{{\mu }_b}\pm\delta \mu )\theta (
     {{\mu }_b}\mp\delta \mu -
      {\sqrt{{\bm p^2}+{m^2}}}
      )
+\frac{i}{2}
\int \frac{d^4 p}{(2\pi)^4}\operatorname{Tr}\log G(p)-\frac{\mu_Q^4}{12\pi^2}
+C,\\
G(p)=&\(\begin{array}{cc}
\slashed{p}-m+\gamma_0(\mu_0+\delta\mu\tau_3+\mu_3\sigma_3+\mu_8)
&-\gamma_5\tau_2\sigma_2\Delta\notag \\
\gamma_5\tau_2\sigma_2\Delta&\slashed{p}-m
-\gamma_0(\mu_0+\delta\mu\tau_3+\mu_3\sigma_3+\mu_8)
\end{array}\), \notag
\end{align}
where we added the contribution to the free energy from the electrons
with chemical potential $\mu_e=-\mu_Q$.  The constant $C$ is fixed
such that $\Omega_{\mathrm{2SC}}(\Delta=0)$ gives the free energy of
the non-interacting system of electrons and quarks.

In calculation the free energy we use the identities~\cite{Bedaque:1999nu}
\begin{align}
\operatorname{Tr}\log\(\begin{array}{cc}A&B\\C&D\end{array}\)
=&\operatorname{Tr}\log\(-B C+BDB^{-1}A\),\\
\int \frac{d^4 p}{(2\pi)^4}\log\[(p_0(1+i\delta)+\delta\mu+\mu_3)^2-
(\Delta^2+\epsilon_{\pm}^2)\]
=&i\int \frac{{d^3}p}{{{(2\multsp \pi )}^3}}
  \[ (\delta \mu
      +{{\mu }_3}+{\sqrt{{{\Delta }^2}+
          {{{{\epsilon }_{\pm }}}^2}}})\theta (\delta
        \mu +{{\mu }_3}
       +{\sqrt{{{\Delta }^2}+{{{{\epsilon }_{\pm }}}^2}}}
        )\notag\right.\\
& \left.
+(\delta \mu
      +{{\mu }_3}-{\sqrt{{{\Delta }^2}+
          {{{{\epsilon }_{\pm }}}^2}}})\theta (\delta
        \mu +{{\mu }_3}
       -{\sqrt{{{\Delta }^2}+{{{{\epsilon }_{\pm }}}^2}}}
        )
\]+ \kappa\notag,
\end{align}
where $\kappa$ 
is a regularization dependent constant that does not depend on the 
BCS gap $\Delta$. For example $\kappa=0$ in dimensional regularization. 
In other regularization schemes such as momentum cutoff, this constant can 
be absorbed into the over all constant $C$ defined in 
Eqs.~(\ref{freeenergy}) and~(\ref{omegadelta2sc}).   
 A straightforward calculation gives
\begin{align}\label{omegadelta2sc}
\Omega_{\mathrm{2SC}}(\Delta)=&-\frac{\mu_Q^4}{12\pi^2}+
\frac{{{\Delta }^2}}{4\multsp g}
  -2\int \frac{{d^3}p}{{{(2\multsp \pi )}^3}}
   {\sum_\pm }\[(\delta \mu
      +{{\mu }_3}+{\sqrt{{{\Delta }^2}+
          {{{{\epsilon }_{\pm }}}^2}}})\theta (\delta
        \mu +{{\mu }_3}
       +{\sqrt{{{\Delta }^2}+{{{{\epsilon }_{\pm }}}^2}}}
        )\right.\\
&\left.+(\delta \mu +{{\mu }_3}
         -{\sqrt{{{\Delta }^2}+
              {{{{\epsilon }_{\pm }}}^2}}})\theta
            (\delta \mu +{{\mu }_3}
           -{\sqrt{{{\Delta }^2}+
                {{{{\epsilon }_{\pm }}}^2}}})\right.\notag\\
&\left.+(\delta \mu -{{\mu }_3}
             +{\sqrt{{{\Delta }^2}+
                  {{{{\epsilon }_{\pm }}}^2}}})\theta
                (\delta \mu -{{\mu }_3}
               +{\sqrt{{{\Delta }^2}+
                    {{{{\epsilon }_{\pm }}}^2}}})\right.\notag\\
&\left.+(\delta \mu -{{\mu }_3}
                 -{\sqrt{{{\Delta }^2}+
                    {{{{\epsilon }_{\pm }}}^2}}})\theta
                    (\delta \mu -{{\mu }_3}
                   -{\sqrt{{{\Delta }^2}+
                    {{{{\epsilon }_{\pm }}}^2}}})\]+C\notag,\\
\epsilon_{\pm}(p)=&\sqrt{\bm p^2+m^2}\pm(\mu_0+\mu_8). \notag
\end{align}
The momentum integrals are evaluated with a cutoff $\Lambda$.  The
coupling $g(\Lambda)$ is determined by requiring a BCS gap
$\Delta_0=100$ MeV at $\mu=350$ MeV and $\mu_Q=0=\mu_3=\mu_8$.  The
light quark masses were assumed to be small compared to $\mu$ and we set them equal to
$5$ MeV.
\begin{figure}[ht]
\begin{center}
\includegraphics[width=.8\textwidth]{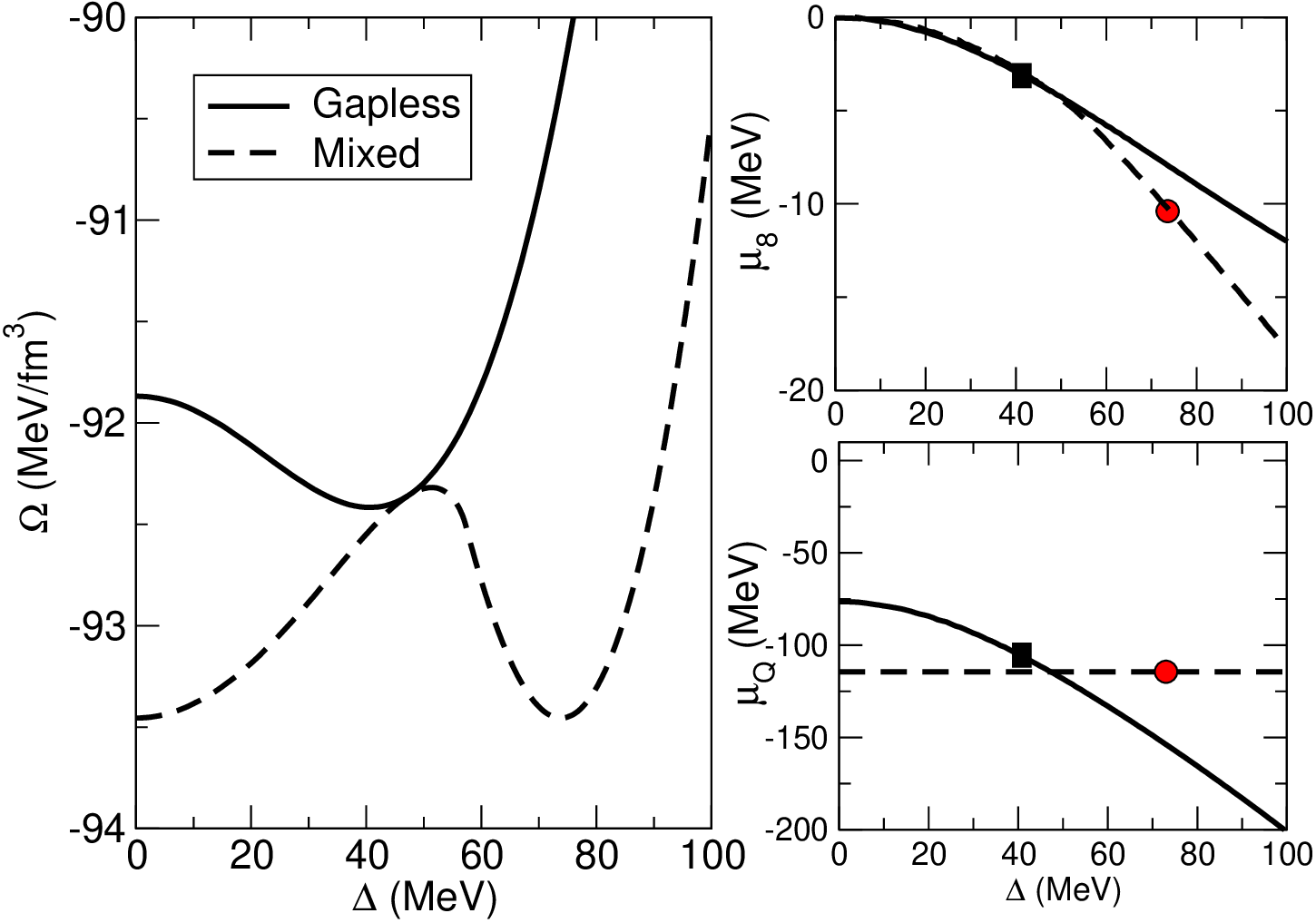}
\end{center}
%\centerline{\epsfxsize 14cm \epsffile{2sc_free.eps}}
\caption{In the left panel, 
free energy curves for the gapless (solid) and the mixed phases
(dashed) as a function of $\Delta$ for $\mu=350 $ MeV are shown. The gapless
free energy is characterized by the imposition of local electric and
color charge neutrality while in the mixed phase electric charge
neutrality is imposed globally.  These requirements induce chemical
potentials $\mu_Q$ and $\mu_8$ and are shown in the right panels.  The
filled dots and rectangles indicate the locations of the mixed and the
gapless ground states.}
\label{omegas}
\end{figure}

Bulk matter is neutral with respect to electric and color
charge.  In what follows we  investigate both the homogeneous and
the heterogeneous phases.  For the former, electric and color charge
neutrality is a local condition. In this case, for a given baryon
chemical potential $\mu$, we require that
\begin{align}
-\frac{\partial\Omega_{\mathrm{2SC}}}{\partial\mu_Q}=0,\quad
-\frac{\partial \Omega_{\mathrm{2SC}}}{\partial \mu_3}=0=
-\frac{\partial\Omega_{\mathrm{2SC}}}{\partial\mu_8}\,.
\end{align}
These sets of equations determine $\mu_Q$, $\mu_3$, $\mu_8$ at any
value of the gap $\Delta$.  Simultaneously solving the above set of
equations together with the gap equation $\partial
\Omega_{\mathrm{2SC}}/\partial \Delta=0$ determines the gapless
states. Denoting the solutions as $\tilde{\mu}_Q$, $\tilde{\mu}_3$,
$\tilde{\mu}_8$ and $ \tilde{\Delta}$, the free energy of the gapless
phase is
\begin{equation}
\Omega_{\mathrm{gapless}}(\mu) \equiv \Omega_{\mathrm{2SC}}(\mu,\tilde{\mu}_Q, \tilde{\mu}_3, \tilde{\mu}_8,\tilde{\Delta}) \,.
\end{equation}

To construct the mixed phase comprising of normal and superconducting
states, electric charge neutrality is imposed as a global
constraint. In principle we could have imposed color neutrality as a
global constraint in the heterogeneous phase. However, we expect
the color Debye screening length, $\lambda_{\mathrm{Debye-color}}
 \simeq 1/\mu $, to be short and comparable to the
inter-particle distance in strong coupling. Under these conditions,
color neutrality is essentially a local constraint. For normal and
superconducting phases to co-exist they must satisfy the Gibbs
criteria: for a given $\mu$, it requires equality of pressures at
equal electric chemical potential.  This results in the following
condition
\begin{align}
\Omega_{\mathrm{2SC}}(\Delta=0,\overline\mu_Q)=\Omega_{\mathrm{2SC}}
(\Delta_{\mathrm{BCS}},\overline\mu_Q) \,,
\label{gibbs}
\end{align}
which in turn uniquely determines $\overline\mu_Q$ corresponding to
the mixed phase. Global charge neutrality follows as long as the
normal and superconducting phases have opposite electric charge. The
volume fraction of the superconducting phase, which we call $\chi$ is 
determined by the global constraint
\begin{align}
\chi~Q_\mathrm{SF}+(1-\chi)~Q_\mathrm{Normal}=&0 , \\
\mathrm{where} \quad 
Q_{\mathrm{SF}}=-\frac{\partial \Omega_{\mathrm{2SC}} (\Delta_{\mathrm{BCS}},\mu_Q)} {\partial\mu_Q}\Big|_{\overline\mu_Q},&\ \ 
Q_{\mathrm{Normal}}=-\frac{\partial \Omega_{\mathrm{2SC}}
(\Delta=0,\mu_Q)}{\partial\mu_Q}\Big|_{\overline\mu_Q},\notag
\end{align}
are the electric charge densities of the superconducting and normal
phases respectively.  

In Fig.~\ref{omegas}, we show the free energy $\Omega_{\mathrm{2SC}}$
as a function of $\Delta$ for $\mu=350$ MeV.  The solid curve is for
the case where electric and color charge neutrality conditions are
imposed locally.  The minimum seen at $\Delta \simeq 40$ MeV
corresponds to the gapless superconducting state described earlier.
The dashed curve shows the free energy corresponding to the mixed
phase.  Here we have imposed only local color neutrality and
determined the electric chemical potential using
Eq.~(\ref{gibbs}). Accordingly, it shows the existence of two
degenerate minima corresponding to the normal and BCS superconducting
states. The variation of the electric and color chemical potentials
are also shown in Fig.~\ref{omegas}.  We find that $\tilde{\mu}_3=0$
and $ |\tilde{\mu}_8|\ll|\overline{\mu}_Q|< |\mu_B|$. The electric
charge chemical potential in the mixed phase is large. For our choice of model parameters and for $\mu$ in
the interval $300-500$ MeV we find that the ratio
$\Delta_0/\delta\mu \simeq 1.3$.  This is consistent with, but smaller than, the weak-coupling leading order prediction of $\Delta_0/\delta\mu = \sqrt{2}$.
On the other hand, $\tilde{\mu}_8$ is
small and proportional to $\Delta_0^2/\mu$ since it is only through pairing
that color neutrality is upset. 

%====================================
\section{Coulomb and Surface Energy in the Mixed Phase} 
\label{mixed}
%====================================
The mixed phase free energy in Fig.~\ref{omegas} ignores finite size
contributions arising due to Coulomb and surface energies associated
with phase separation. If these effects are negligible, the results
clearly indicate that the mixed phase is favored over the gapless
phase. However, given the small difference in free energy between the
gapless and mixed phases including these corrections becomes
necessary. In Sec.~\ref{surface} we calculate the surface energy in
the mixed phase. For now, we treat the surface tension as a parameter
and study how it influences the free energy of the mixed phase.

The mixed phase can be subdivided into electrically neutral unit cells
called Wigner-Seitz cells, as in the analysis of the inner crust of a
neutron star where droplets of charged nuclear matter coexist with a
negatively charged fluid of neutrons and
electrons~\cite{Negele:1973vb}.  In the present context, each
Wigner-Seitz cell will contain some positively charged superconducting
quark matter and some negatively charged normal quark matter.
Although at low temperature these unit cells form a Coulomb lattice,
the interaction between adjacent cells can be neglected compared to
the surface and Coulomb energy of each cell.  In this Wigner-Seitz
approximation, the surface and Coulomb energy per unit volume are
fairly straightforward to calculate if the charge density is spatially
uniform in each phase and if the surface thickness is small compared
to the spatial extent of the Wigner-Seitz cell. We will examine both
these requirements in greater detail in Sec.~\ref{surface}.  For now
we shall assume that these requirements are satisfied. In this case,
the surface and Coulomb energies depend in general on the geometry and
are given by~\cite{Ravenhall:1983uh}
\begin{eqnarray}
E^S &=& \frac{d~x~\sigma}{r_0} \,, \label{surface_ener}\\
E^C &=& 2\pi~\alpha_{\rm em}f_d(x)~x~(\Delta Q)^2~r_0^2 \,, 
\label{coulomb_ener}
\end{eqnarray}
where $d$ is the dimensionality of the structure ($d=1,2,$ and $3$
correspond to Wigner-Seitz cells describing slab, rod and droplet
configurations, respectively), $\sigma$ is the surface tension,
$\Delta Q=Q_{\rm SF} -Q_{\rm Normal}$ is the charge density contrast
between the two phases and $\alpha_{\rm em}=1/137$ is the fine
structure constant. The other factors appearing in
Eqs.~(\ref{surface_ener}) and (\ref{coulomb_ener}) are: $x$, the fraction
of the rarer phase which is equal to $\chi$ for $\chi \leqslant 0.5$
and $1-\chi$ for $0.5 < \chi \leqslant 1$; $r_0$, the
radius of the rarer phase (radius of drops or rods and half-thickness
of slabs); and $f_d(x)$, the geometrical factor that arises in the
calculation of the Coulomb energy which can be written as
\begin{equation}
f_d(x)=\frac{1}{d+2}~\left(\frac{2-d~x^{1-2/d}}{d-2} + x\right) \ .
\end{equation}
The first step in the calculation is to evaluate 
$r_0$ by minimizing the sum of $E^C$ and $E^S$. The result is
\begin{equation}
r_0 = \left[\frac{d~\sigma}{4\pi~\alpha_{\rm
em}f_d(x)~(\Delta Q)^2}\right]^{1/3} \,.
\label{radius}
\end{equation}
We then use this value of $r_0$ in
Eqs.~(\ref{surface_ener}) and (\ref{coulomb_ener}) to evaluate the surface
and Coulomb energy cost per unit volume
\begin{equation}
E^S+E^C = \frac{3}{2} \left(4\pi~\alpha_{em}~d^2~f_d(x)~x^2\right)^{1/3}
~(\Delta Q)^{2/3}~\sigma^{2/3} \,.
\label{sandccost}
\end{equation}
This equation allows us to include the free energy cost associated
with the Coulomb field and the surface. After calculating the surface
tension in Sec.~\ref{surface} we will return to an analysis of the free
energy competition between gapless and mixed phases in
Sec.~\ref{phase}.
%====================================
\section{Surface Tension} 
\label{surface}
%====================================
\begin{figure}[ht]
\begin{center}
\includegraphics[width=0.7\textwidth]{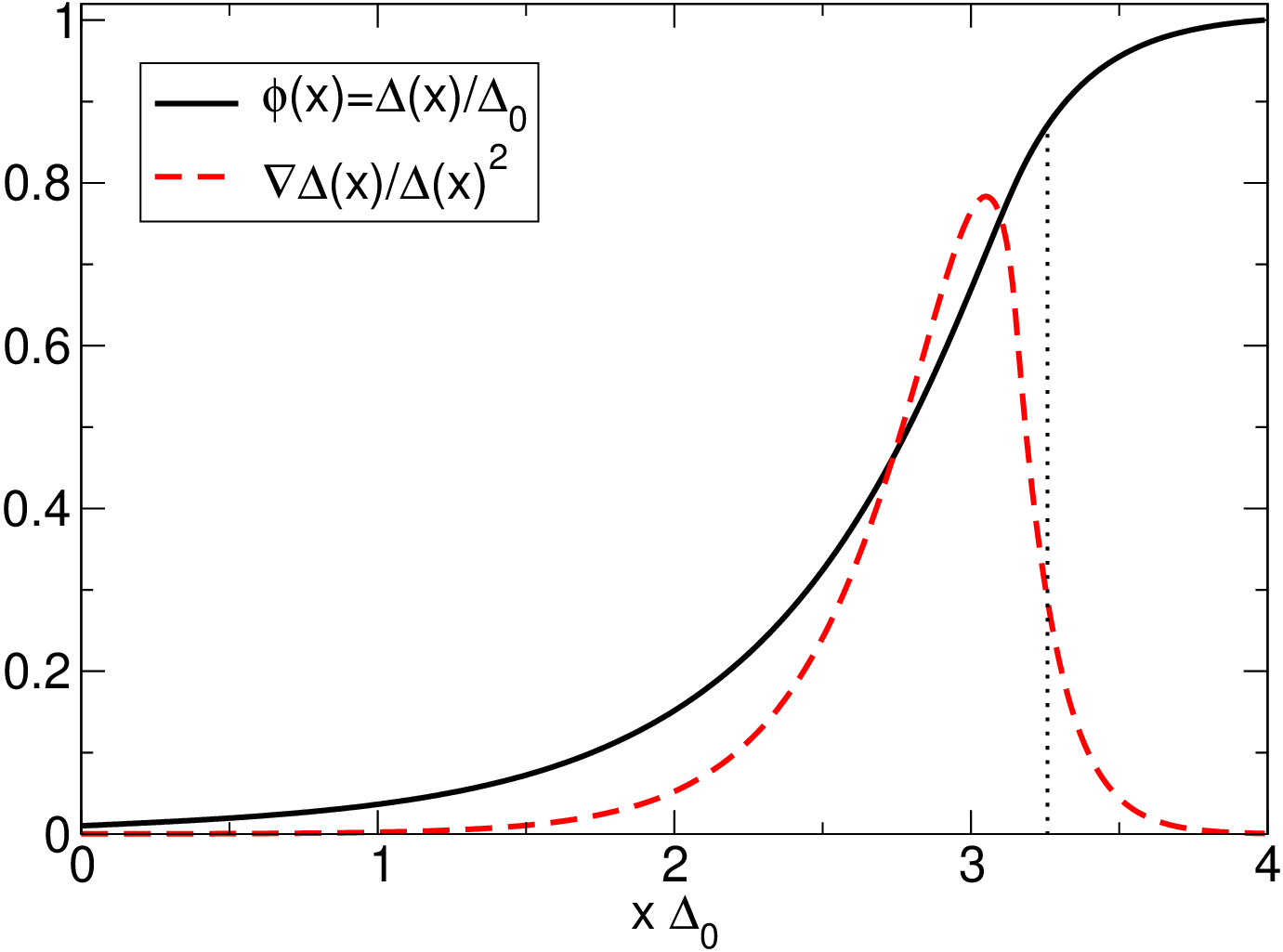}
%\centerline{\epsfxsize 10cm \epsffile{profile.eps}}
\end{center}
\caption{The profile of $\Delta(x)$ at the interface for $\mu=400$
MeV. The solid curve shows the variation of the dimensionless quantity
$\phi(x)=\Delta(x)/\Delta_0$ as function of the $\bar{x}=x
\Delta_0$. The dotted-line corresponds to the region where the leading
order gradient energy is less than zero when $\delta\mu$ corrections
from Ref.~\cite{Huang:2004bg} are included. The dashed curve shows the variation of  $\nabla\Delta(x)/\Delta(x)^2$. }
\label{interface}
\end{figure}
We will now calculate the surface energy associated with the
normal-superconductor interface. This will allow us to determine if
the mixed phase is favored over the gapless state.  To accomplish this
we will need to allow for spatial variations of $\Delta$ and calculate
the gradient contribution to the free energy.  Near the critical
temperature one can employ Landau-Ginsburg theory to obtain the
leading gradient contribution to the free energy. In this case, the
gradient expansion (in powers of $|\nabla~\Delta(x)|^2$) is well
defined. Bailin and Love have derived an expression for the
gradient contribution to the free energy for a relativistic
superconductor near the critical temperature \cite{Bailin:1984bm}. In the context of dense quark matter, a discussion of the Landau-Ginsburg theory for two and three flavor quark matter can be found in Refs.~\cite{Iida:2000ha,Voskresensky:2004jp}. For
temperatures that are small compared to the critical temperature, the
gradient expansion (in powers of $|\nabla~\Delta(x)|^2$) has poor
convergence properties and one must in principle include gradients to
all orders. However, if the spatial variations of $\Delta$ occur on a
length scale that is numerically large compared to $\Delta^{-1}$, it
is a good approximation to retain only the leading term. We shall
assume that the thickness of the normal-superconductor interface is
larger than $\Delta^{-1}$ and retain only the leading contribution
from the gradient term. In this case, the $\Delta$ dependent
contribution to the free energy of the 2SC phase, including the
gradient term, is given by
\begin{align}
\label{landauginzburg}
\mathcal{F}(\Delta)_{\mathrm{2SC}} =& \Omega_{\mathrm{2SC}} +
\kappa_{\mathrm{2SC}}^{(2)} |\nabla~\Delta(x)|^2 
+ {\mathrm O}(|\nabla~\Delta(x)|^4/\Delta^4) 
 \,.
\end{align} 
The coefficient of the gradient term ($\kappa_{\mathrm{2SC}}$) for the
2SC phase, which is related to the Meissner mass of the gluons, at
zero temperature and for the case where $\delta\mu=0$ has been
calculated in earlier work by Rischke \cite{Rischke:2000qz}.
Recently, Huang and Shovkovy have investigated how $\delta\mu$ affects
the Meissner masses (which as discussed earlier, in turn determines
$\kappa_{\mathrm{2SC}}$) \cite{Huang:2004bg}. They find that the
Meissner mass of the eight gluon (assuming that the condensate aligns
in the 3 direction) remains unaffected by $\delta\mu$ in the BCS phase
i.e, for $\delta\mu \le \Delta_0$. On the other hand, they find the
Meissner masses of gluons with color 4,5,6,7 (which are degenerate)
decrease with increasing $\delta\mu$ and become negative when
$\delta\mu = \Delta_0/\sqrt{2}$.  Surprisingly, these gluon masses
become negative even in the BCS phase, i.e, for $\Delta_0/\sqrt{2} \le
\delta\mu \le \Delta_0$ . It is presently not clear how this
instability is resolved.  In this work, where we have employed an Nambu-Jona-Lasino (NJL)
model description of the interaction between quarks, the gluons are
introduced as fictitious gauge fields solely for the purpose of
determining the coefficient $\kappa_{\mathrm{2SC}}$ in the effective
theory for the real field $\Delta$ \cite{Son:1999cm}.  By minimal
coupling between the field $\Delta$ and the gluons, the relation
between $\kappa_{\mathrm{2SC}}$ and the Meissner mass ($m_g^2~\simeq~ \alpha_s^2
\mu^2$) of the gluons is obtained by matching to be 
$\kappa_{\mathrm{2SC}}^{(2)} \propto m_g^2/\alpha_s^2\Delta_0^2$, where $\alpha_s$
is the strong coupling constant.  

In our calculation, the coefficient $\kappa_{\mathrm{2SC}}^{(2)}$ is
determined by matching the mass of the fourth and eight gluon in the
effective theory to the microscopic calculation.  We note that
$\kappa_{\mathrm{2SC}}^{(2)}$ could depend on $\Delta(x)$, since the
low energy effective theory for the real field $\Delta(x)$ does not
have a well defined expansion parameter. Hence, in general
\begin{equation}
\label{kappaexp}
\kappa_{\mathrm{2SC}}^{(2)}= \sum_{n=0,\infty}~a_n
~\left(\frac{\Delta(x)}{\Delta_0}\right)^{2n}\,,
\end{equation}
where $a_n$ are constants that remain to be determined by matching
conditions.  We make the simplifying assumption made in
Ref.~\cite{Rischke:2000qz} and retain only the leading ($a_0$) and
next to leading ($a_1$) order terms. In this case, using the Meissner
masses calculated in Ref.~\cite{Huang:2004bg} and for $\Delta_0 \ge \delta \mu $ we obtain
\begin{equation}
\label{kappat0}
\kappa_{\mathrm{2SC}}^{(2)}= \frac{\bar{\mu}^2}{3\pi^2~\Delta_0^2}~
\left[(1-2\frac{\delta\mu^2}{\Delta_0^2})
-\frac{1}{2}~(1-4\frac{\delta\mu^2}{\Delta_0^2})
\frac{\Delta(x)^2}{\Delta_0^2}\right]\,,
\end{equation}
where $\bar{\mu}=\mu+\mu_Q/6+\mu_8$ is the common chemical potential
of the quarks participating in pairing and $\delta\mu=\mu_Q/2$. 
From Eq.~\ref{kappat0} we
see that for $\Delta_0/\delta\mu \le \sqrt{2}$, the coefficient $a_0
\le 0$ while $a_1 $ remains positive in the mixed phase where
$\Delta_0/\delta\mu \ge 1$. In weak coupling and to leading order in the $\Delta_0$ and $\delta\mu$, the mixed phase occurs precisely at 
 $\Delta_0/\delta\mu = \sqrt{2}$. In our
calculation of the mixed phase free energy in Sec.~\ref{free}
we found that $\Delta_0/\delta\mu \simeq 1.3$ for  $\mu$ in the interval $300-500$ MeV.  This result, which implies that $\Delta_0/\delta\mu < \sqrt{2}$  is {\it not} generic to the mixed phase. For other values of the coupling we indeed find that $\Delta_0/\delta\mu \ge \sqrt{2}$  and both $a_0$ and $a_1$ are positive. Nonetheless, even when $a_0 \le 0$, for  $\Delta(x)/\Delta_0 \simeq 1$,
the $a_1$ term ensures that the gradient contribution is positive and
is therefore crucial for the stability of the BCS phase component of
the mixed phase. In the vicinity of the BCS phase, i.e, when
$\Delta(x)/\Delta_0 \simeq 1$, the gradient contribution is
independent of $\delta\mu$ and $\kappa_{\mathrm{2SC}}^{(2)} =
\bar{\mu}^2/6\pi^2\Delta_0^2$. We note that had we retained only the $a_0$ contribution to $\kappa_{\mathrm{2SC}}^{(2)}$ and matched to 
the mass of the eight-gluon (arguing that in our NJL model treatment 
only the generators $T_3$ and $T_8$ are relevant) we would still have obtained  $\kappa_{\mathrm{2SC}}^{(2)} =\bar{\mu}^2/6\pi^2\Delta_0^2$.  A non-zero $\delta\mu$ in general acts
to reduce the energy cost associated with the gradient term when
$\Delta(x)/\Delta_0$ decreases from unity. Surprisingly, when
$\Delta(x)/\Delta_0 =
\sqrt{2(2(\delta\mu^2/\Delta_0^2)-1)/(4(\delta\mu^2/\Delta_0^2)-1)}$, the
gradient energy goes to zero. This does not necessarily mean that the
interface is unstable since we have not investigated the higher order
gradient terms.  However, this trend suggests that we can obtain an
upper bound on the surface tension by using the (largest) value of
$\kappa_{\mathrm{2SC}}^{(2)}=\bar{\mu}^2/6\pi^2\Delta_0^2$ across the
whole interface. We note in passing that for temperatures near $T_C$, where
$T_C$ is the BCS critical temperature, Bailin and Love
\cite{Bailin:1984bm} find that
\begin{equation}
\kappa_{\mathrm{2SC}}^{(2)}=\frac{7\zeta(3)}{24\pi^4}~
\frac{\mu^2}{\alpha^2~\Delta_0^2} \,
\label{teqtc}
\end{equation} 
where $\alpha \simeq 0.57 = k_BT_C/\Delta_0$, $\zeta(3)\simeq
1.202$. Interestingly, the results for $\kappa_{\mathrm{2SC}}^{(2)}$ in
the vicinity of the BCS state at $T=0$ and $T \sim T_C$ are both
parametrically and numerically similar.

The contribution to the free energy which is independent of spatial
derivatives is given by $\Omega_{\mathrm{2SC}}(\Delta)$ and was
derived earlier in Eq.~(\ref{omegadelta2sc}). In general, both the
gradient term and $\Omega_{\mathrm{2SC}}$ are functions of all three
chemical potentials, namely: $\mu$, $\mu_Q$ and $\mu_8$. While we
retain this full dependence in $\Omega_{\mathrm{2SC}}$ we use the
approximate form for $\kappa_{\mathrm{2SC}}^{(2)}$ valid when
$\Delta(x)\simeq \Delta_0$. To construct the spatial profile of
$\Delta$ between the normal and superconducting phases we use the
equation of motion. In one spatial dimension this is given by
\begin{equation} 
\kappa_{\mathrm{2SC}} \frac{d^2\Delta(x)}{dx^2} + 
(\frac{d\Delta(x)}{dx})^2 
\frac{\partial\kappa_{\mathrm{2SC}}}{\partial \Delta} =  
\frac{\partial\Omega_{\mathrm{2SC}}}{\partial \Delta} \,.
\label{profile}
\end{equation} 
We solve this ordinary differential equation subject to the boundary
condition that $\Delta(x=0)=0$ and $\Delta(x)=\Delta_0(\mu)$ at large
$x$, where $\Delta_0(\mu)$ is the gap in the BCS phase. The results
are shown in Fig.~\ref{interface}, where we calculate the interface
for various values of the baryon chemical potential $\mu=400$ MeV. We
find that the profiles for the interface for other values of $\mu$ in
the range $350-450$ MeV show nearly identical behavior when plotted in
terms of the reduced dimensionless units
$\phi(x)=\Delta(x)/\Delta_0(\mu)$ and $\bar{x}=x \Delta_0 (\mu)$.  The
typical length scale for the variation of $\Delta$ is parametrically
similar to but numerically larger than $\Delta_0^{-1}$. We find that a
rough measure of the thickness $t \simeq (1.5-2)  \Delta_0^{-1}$. As
mentioned earlier, due to the lack of a well defined gradient
expansion for effective theory at $T=0$, the validity of the leading
order (in the gradient expansion) results requires that the spatial variation of $\Delta(x)$ be
weak and satisfy the condition $\nabla \Delta/\Delta_0^2 \ll 1$. From
Fig.~\ref{interface}, we see that this condition is satisfied to a
fair degree (see dashed lines). Using these profiles, we have computed
the surface tension by integrating the free energy function given in
Eq.~(\ref{landauginzburg}). The results are presented in the table
below.
\begin{center}
\begin{tabular}{|c|c||c|c||c|c|} \hline
$\mu$ (MeV)  & $\sigma$ (MeV/fm$^2$) & 
$\mu$ (MeV)  & $\sigma$ (MeV/fm$^2$)  & 
$\mu$ (MeV)  & $\sigma$ (MeV/fm$^2$)  \\ \hline 
350 & 1.9  & 375 &  2.6 & 400 & 3.6 \\ \hline
425  & 4.9   & 450 &  6.8   & 475  & 9.4 \\ \hline
\end{tabular}
\end{center}
Our estimation of the coefficient of the kinetic term, which we call
$\kappa_{\mathrm{2SC}}^{(2)}$ ignores several possible
corrections. Including those due to the truncation of the series in
Eq.~\ref{kappaexp}, and the neglect of $\delta\mu$.  As discussed
earlier the effect of $\delta\mu$ is to decrease the gradient
contribution, consequently we can expect these corrections to lower
our estimate of the surface tension. In Fig.~\ref{interface}, the
vertical dotted-line indicates the region where the gradient energy is
less than zero when $\delta\mu$ corrections are included. In this
case, the profile would resemble the dotted-line rather than the solid
line and corresponding surface energy is greatly reduced. We have also
examined how an increase in $\kappa_{\mathrm{2SC}}^{(2)}$ would affect
our calculation of the surface tension. We find that these changes are
modest. The surface tension $\sigma$ increased by $40\%$ when
$\kappa_{\mathrm{2SC}}$ was increased by a factor of two.
%-------------------------------
\section{Phase Diagram}
\label{phase} 
%-----------------------------------
\begin{figure}[ht]
\begin{center}
\includegraphics[width=0.9\textwidth]{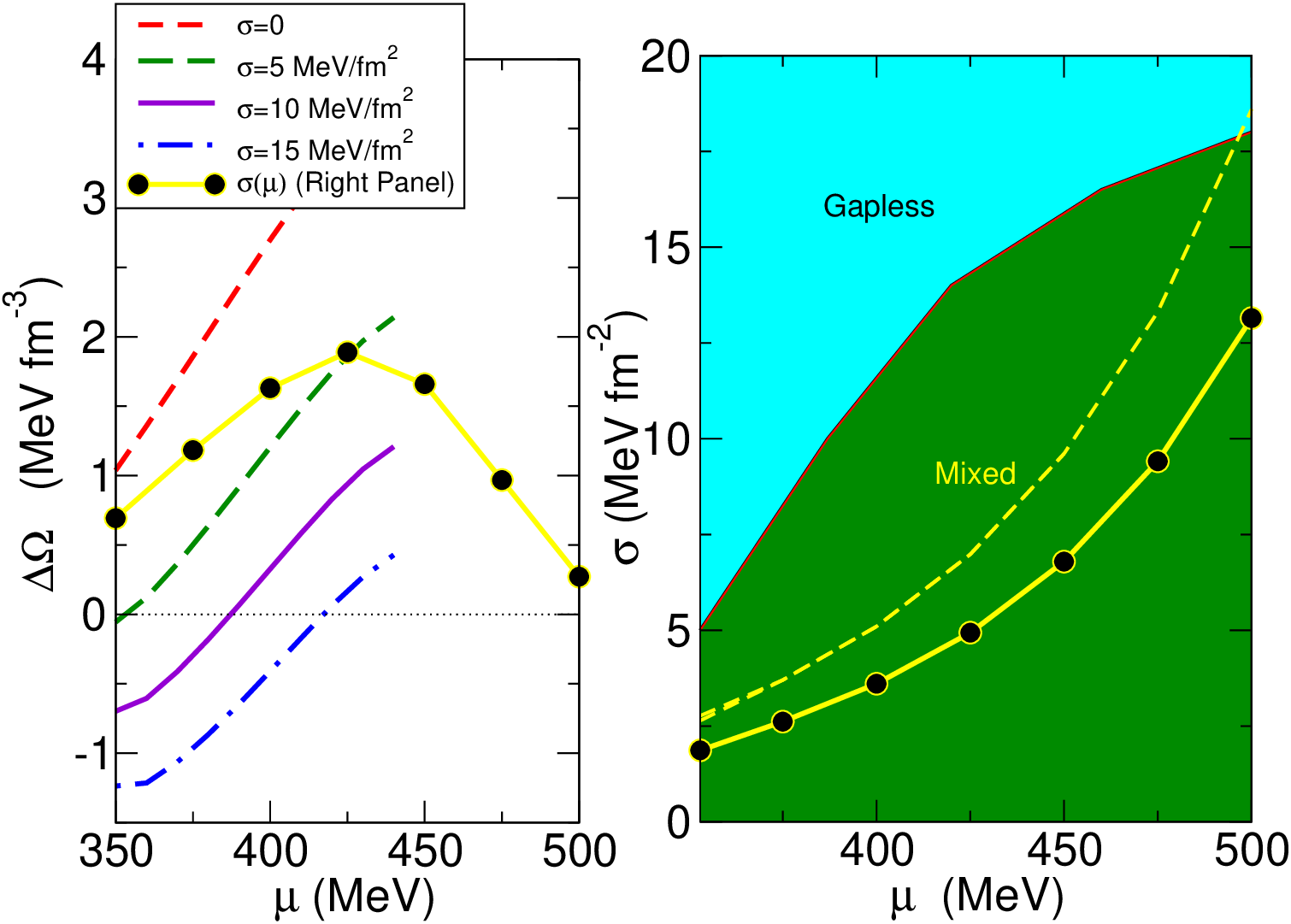}
%\centerline{\epsfxsize 14cm \epsffile{phase_diagram.eps}}
\end{center}
\caption{\protect Left panel: Influence of the surface energy on the
total free energy difference between the gapless and the mixed phases
$\Delta\Omega =\Omega_{\mathrm{gapless}}- \Omega_{\mathrm{mixed}}$ for
various values of the surface tension. The filled circles correspond
to values of $\sigma(\mu)$ shown in the right panel. Right panel: The
phase diagram showing the favored phase as a function of the surface
tension and baryon chemical potential. The filled circles correspond
to the results obtained from calculation of the surface tension
outlined in Sec.~\ref{surface}. The dashed line shows the surface
tension when $\kappa_{\mathrm{2SC}}^{(2)}$ was increased by a factor of
two.}
\label{phasediagram}
\end{figure}

Using the results of the previous sections, we can study the free
energy competition between the gapless and the mixed phases. The free
energy of the mixed phase, including Coulomb and surface contributions
is
\begin{equation} 
\Omega_{\mathrm{mixed}}(\mu) = \Omega_{\mathrm{2SC}}(\mu,\overline{\mu}_Q) +
E^S + E^C\,,
\end{equation}
where the surface and the Coulomb contributions were defined in
Eq.~(\ref{sandccost}) and $\overline{\mu}_Q$ is defined through
Eq.~(\ref{gibbs}).  The left panel of the Fig.~\ref{phasediagram}
shows the free energy difference $\Delta\Omega
=\Omega_{\mathrm{gapless}}- \Omega_{\mathrm{mixed}}$ for various
values of the surface tension in the mixed phase. The filled circles
correspond to the values of the surface tension computed in
Sec.~\ref{surface}. We have used this information in constructing the
"phase diagram" as a function of $\mu$ and $\sigma$, and is shown in
the right panel of the figure. As expected, the gapless homogeneous
phase is favored when $\sigma$ is large and the mixed phase is favored
for small $\sigma$. The surface tension computed in the previous
section is also shown (filled circles). Their low values at small
$\mu$ indicate that the mixed phase is favored at low density. At
first, with increasing density the free energy difference
($\Delta\Omega$) increases, resulting in a robust mixed phase. For
$\mu \gsim 450$ MeV the free energy difference decreases, indicating a
possible first-order transition to the gapless phase at high
density. The dashed dashed-curve on the right panel shows the surface
tension for the case where we artificially increased
$\kappa_{\mathrm{2SC}}^{(2)}$ by a factor of two. In this case the
trends are similar, the mixed phase continues to be favored at low
density and a first order transition to the gapless phase occurs at
$\mu \simeq 500$ MeV. To eliminate the mixed phase
$\kappa_{\mathrm{2SC}}^{(2)}$ has to be at least on order of magnitude
larger than the one used in this work.

We now examine if the assumptions made earlier in constructing the
mixed phase are satisfied for the values of the surface tension
calculated in Sec.~\ref{surface}.  The Debye screening length in the
normal and superconducting phases are easily computed by using the
relation $\lambda_D^{-2} =-4\pi~\alpha_{\mathrm{em}}~ \partial^2
\Omega_{\mathrm{2SC}}/\partial \mu_Q^2$. For the individual phases to
have uniform charge distributions, their spatial extents in the mixed
phase must be small compared to the Debye screening length.  The table
below provides a comparison between these length scales at various
values of the baryon chemical potential. The results are shown for the
mixed phase with $d=3$, corresponding to a spherical geometry for the
Wigner-Seitz.  In the table, $\Delta_{\mathrm{BCS}}$, $\chi$,
$R_{\mathrm{D}}$ and $R_{\mathrm{ws}}$ correspond to the gap in the
BCS phase, the volume fraction of the BCS phase, the droplet radius
and the radius of the Wigner-Seitz cell respectively. Further, we have
also implicitly assumed that the interface thickness is small compared
to $R_{\mathrm{D}}$ and $R_{\mathrm{ws}}$. The thickness parameter
$t$, defined to be the distance over which the gap roughly decreases
by a factor of $e\simeq 2.718$, is also given in the table.
\begin{center}
\begin{tabular}{|c|c|c|c|c|c|c|c|} \hline
$\mu$ (MeV) & $\Delta_{\mathrm{BCS}}$ (MeV) & $\chi$ & $R_{\mathrm{D}}$ (fm) & 
$R_{\mathrm{ws}}$ (fm) &  $\lambda_D^{\mathrm{sc}}$ (fm) & 
$\lambda_D^{\mathrm{normal}}$ (fm) &  $t$ (fm) \\ \hline 
350 & 74 & 0.4 & 6 & 8 & 7 & 5 &  4  \\ \hline 
400 & 106 & 0.7 & 4 & 6 & 6 & 4.5 & 3   \\ \hline 
450 & 140 & 0.8 & 3 & 5 & 5 & 4 & 2 \\ \hline 
500 & 174 & 0.9 & 3 & 5 & 5 & 4 & 2 \\ \hline 
\end{tabular}
\end{center}
The length scales shown in the above table are just barely compatible
with our earlier assumptions. In particular, the Debye screening
length and the spatial extents are of similar size, the charge
distribution in the mixed phase could differ from the simple profiles
considered here.  We also note that the thickness is not small
compared to the typical size of the droplet or the Wigner-Seitz
cell. We will not consider these finite size corrections in this work.
These corrections generically tend to decrease the Coulomb energy, increase the surface energy and penalize the mixed phase when the volume fraction $\chi$ is near zero or unity \cite{Norsen:2000wb,Voskresensky:2002hu}. These finite size effects in the mixed phase clearly warrant further study. 

\begin{figure}[h]
\begin{center}
\includegraphics[width=0.6\textwidth]{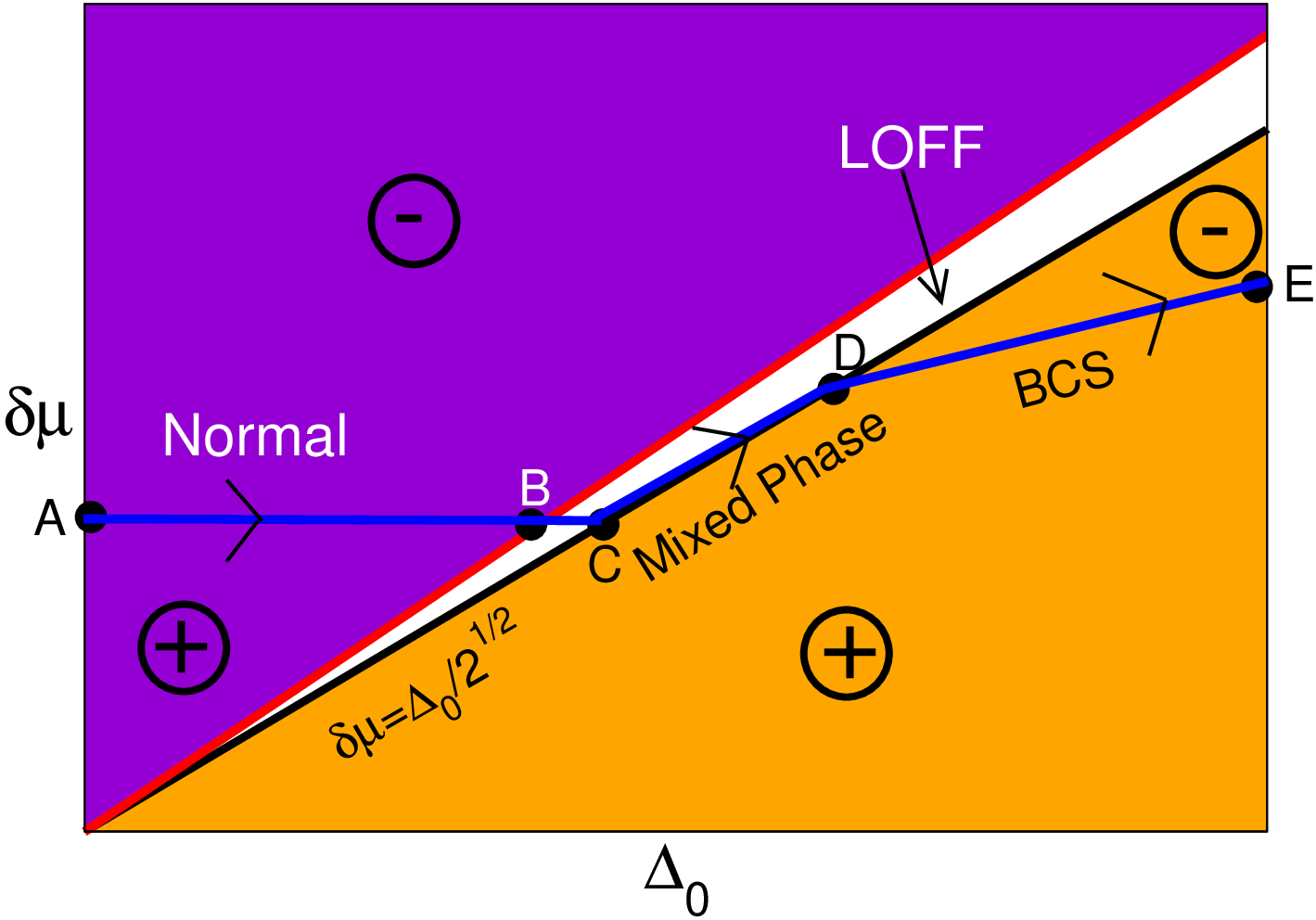}
%\centerline{\epsfxsize 10cm \epsffile{profile.eps}}
\end{center}
\caption{A schematic phase diagram at mixed $\mu$. The BCS gap $\Delta_0$ (or equivalently the effective four-fermion coupling constant) is plotted along the x-axis and $\delta\mu=\mu_Q/2$ is the y-axis. The line $\delta\mu=\Delta_0/\sqrt{2}$ separates the regions where the BCS state (lower-right)  and the normal states (upper-left) are favored. The charge-neutral ground state follows the trajectory shown by the thick lines with arrows. The wedged region  between the lines corresponding to 
$\delta\mu=\Delta_0/\sqrt{2}$ and $\delta\mu=0.75\Delta_0$ indicates the region were the LOFF phase may be favored\cite{Alford:2000ze}.}
\label{schema}
\end{figure}

The results of this work combined with the finding of earlier investigations in Refs.~\cite{Alford:2000ze,Alford:2002kj,Steiner:2002gx,Neumann:2002jm} of the charge neutral two flavor quark matter suggest a schematic  phase diagram of the form shown in Fig.~\ref{schema}. The figure depicts the phase structure at fixed $\mu$ . The BCS gap $\Delta_0$ or equivalently the effective four-fermion coupling increases along the x-axis  and $\delta\mu=\mu_Q/2$ increases along the y-axis.  The line $\delta\mu=\Delta/\sqrt{2}$ separates the regions where the BCS state (lower-right)  and the normal states (upper-left) are favored (As discussed earlier, this is true only in weak coupling. In strong coupling, $\delta\mu$ lies in the vicinity of $\Delta_0/\sqrt{2}$). The charge-neutral ground state follows the trajectory shown by the thick lines with arrows. In each phase, regions above this line correspond to net positive charge and regions below to net negative charge. For small coupling ($\Delta_0$), the homogeneous normal phase is preferred, and is represented by the line from point A to B. At intermediate coupling, the mixed phase containing an admixture of negatively charged normal phase coexisting with  positively charged BCS phase is favored. This is shown by the line from point C to D. At large coupling, the homogenous BCS or 2SC phase prevails and is depicted by the line from point D to E.  The wedged region  between the lines corresponding to 
$\delta\mu=\Delta/\sqrt{2}$ and $\delta\mu=0.75\Delta$ indicates the region were the LOFF phase may be favored\cite{Alford:2000ze}. It remains to be seen how this phase could exist as a charge-neutral state. If the charge density in the LOFF phase resembles the normal phase, the figure supports the existence of a region between point B and C where the LOFF phase could be stable and electrically neutral. It also points to 
the possibility of a mixed phase of LOFF and BCS. Investigation of the LOFF phase in the charge neutral phase diagram is clearly warranted but is beyond the scope of this article.   

\section{Discussion}
\label{discuss}
Based on our calculation of the surface tension between normal and
superconducting 2-flavor quark matter we have shown that a
heterogeneous mixed phase is likely to be the ground state of quark
matter at low density. The stress induced on the bulk system by the
requirement of charge neutrality is resolved by phase separation. The
other possibility, which is the homogeneous gapless phase is energetically disfavored at low density. As depicted in Fig.~\ref{schema}, at small coupling we can expect the charge neutral 
ground state to be be in the normal phase. At intermediate coupling, the heterogeneous mixed phase is favored and large coupling a uniform BCS like 
2SC phase is the ground state.   

While we expect our qualitative results to be fairly robust, there are
three important caveats to our quantitative findings that we shall now
discuss. First of these caveats is related to the earlier observation
that the surface thickness $t\simeq 1-3 $ fm, the typical physical
sizes of the charged regions $R_{\mathrm{D}} \simeq = 2-6$ fm and the
Debye screening length $\lambda_D \simeq 5-8$ fm are not well
separated. This is likely to alter the simple structure of the mixed
phase considered in this work. The charge, baryon density and the gap
will vary smoothly over most of the Wigner-Seitz cell. These finite
size corrections will need to be examined before one can draw firm
conclusions regarding the phase competition (for a discussion of these
effects in the context of a nuclear-kaon mixed phase see
Ref.~\cite{Norsen:2000wb}). Second, we have assumed that the weak
coupling BCS theory holds even when $\Delta/\mu \simeq 1/3$. In the
BCS limit $\Delta \ll \mu$, neither the gapless nor the mixed phase
could exist because for 2-flavor quark matter the electric charge
chemical potential is proportional to the baryon chemical
potential. In the strong coupling regime it is difficult to know a
priori how other contributions to the free energy would alter this
competition.  Finally, we have only included the leading order
gradient contribution in the effective theory for $\Delta$. Although
our profiles indicate that spatial variation of $\Delta(x)$ is mild
(From Fig.~\ref{interface}, $|\nabla \Delta(x)|^2/\Delta_0^2 \lsim
1$), it is a priori not possible to assess how the higher order spatial
derivatives of $\Delta(x)$ in the free energy would change our
estimate of the surface tension.  As could corrections to the
coefficient $\kappa_{\mathrm{2SC}}^{(2)}$ arising from strong coupling and the lack of well defined expansion parameter for the 
effective theory at $T=0$. Additionally, corrections due to electric and color charge chemical potentials are important \cite{Huang:2004bg}.  Interestingly, the correction to the gradient
term from the electric charge chemical potentials tends to decrease
the surface energy.  We have shown that drastic changes to
$\kappa_{\mathrm{2SC}}^{(2)}$ (an increase by a factor of 2) resulted
in a surface tension that continued to favor the mixed phase at low
density. In this case, with increasing density, we found that the
gapless phase becomes favored. The location of this first order
transition between the mixed phase and the gapless phase is model
dependent since it is sensitive to the density dependence of the ratio
$\mu/\Delta_0$.

An important consequence of $\delta\mu$ corrections to the kinetic
term concerns stability of the gapless phase. First we note that we
can expect the matching between the kinetic coefficients for the
effective theory for $\Delta(x)$ and the masses of gluons in the
microscopic theory to be valid in the gapless phase. Since
$\delta\mu \ge \Delta_0$ in the gapless phase this implies an
instability of the spatially uniform gapless phase (the leading order
gradient energy is negative). How this instability is resolved is not
clear at this time. It is however clear that some type of
heterogeneity must occur. Likely candidate states include the LOFF
phase and the mixed phase. Pending a detailed investigation of the charge
neutral LOFF state, our results suggest that this instability may be
resolved by the formation of a mixed phase. In our NJL model study, the mixed phase is stable with respect to small gradient perturbations
for all $\Delta_0/\delta\mu \ge 1$. In weak coupling, since  $\Delta_0/\delta\mu = \sqrt{2}$ in the mixed phase, the chromo-magnetic instability discussed in Ref.~\cite{Huang:2004bg}  for gluons 4-7 in QCD does not occur in the mixed phase. In strong coupling, whether this instability persists and if so, the nature of its resolution is not yet clear. 

A heterogeneous quark matter ground state has important consequences
for neutron star physics - were quark matter to exist inside their
cores. The free energy difference between the gapless and mixed phase
is small. Consequently, its effects on neutron star structure will be
small. On the other hand, we can expect important changes to the
transport properties induced by the heterogeneity of the mixed
phase. In this phase, transport of the low energy neutrinos, photons, 
etc., will be dominated by coherent scattering induced by the presence
of superconducting ``bubbles'' of quark matter with typical dimension
of $r_0 \simeq 5 $ fm (for a discussion of how heterogeneity affects
low energy neutrino transport see Ref. \cite{Reddy:1999ad}). Further,
the existence of regions with normal quark matter will affect the
neutrino emissivities - since they are typically large in the normal
phase. Finally, the crystalline structure of the mixed phase could be
relevant to the modeling of glitches observed in some pulsars.

We note that there are other mixed phases that could play a role in
neutron stars. These include the mixed phase between quark matter and
nuclear matter\cite{Glendenning:1992vb}; between color-flavor-locked
matter and nuclear matter\cite{Alford:2001zr}, or superconducting
quark matter and normal quark matter\cite{Neumann:2002jm}. All of
these share several common features: the coexistence of positively and
negatively charged phases separated by an interface; crystalline
structure at low temperature; and coincidentally typical droplet sizes
of the order $5$ fm. However, in these aforementioned examples it is
still unclear if the mixed phase is favored when surface and Coulomb
costs are included. This is because the surface tension is poorly
known. Arguments based on naive dimensional analysis indicate that the
surface energy cost may indeed be too large in these systems
\cite{Heiselberg:1993dx,Alford:2001zr}. The exception could be the mixed phases of normal and superconducting quark matter considered in Ref.~\cite{Neumann:2002jm}. In this case, the surface tension is computable and the physical considerations are very similar to those studied in this work. 

Although quantitative aspects of the analysis presented in this work
pertains to asymmetrical relativistic superconductors, our finding
provide some qualitative insight into the behavior of non-relativistic
superfluid that can be realized in ultra cold fermionic-atom traps being
developed in the laboratory. Although there is no Coulomb cost
associated with the mixed phase in these systems, a surface
energy cost must be overcome when the surface to volume ratio is not
negligible. The methods described in Sec.~\ref{surface} apply in
general to asymmetrical fermion systems and its relevance to
laboratory experiments is currently under investigation.

%%%%%%%%%%%%%%%%%%
%  Acknowledgments
%%%%%%%%%%%%%%%%%%

\vskip0.25in \centerline{\bf Acknowledgments} We would like to thank
Mark Alford, Paulo Bedaque, Heron Caldas, Joe Carlson, Krishna
Rajagopal, Igor Shovkovy and Dam Son for useful discussions. This
research was supported by the Dept. of Energy under contract
W-7405-ENG-36.

%=========== Bibliography ==================
%\bibliographystyle{h-physrev4}
%\bibliography{mixed} 

\end{document}